\documentclass[12pt]{article}

\begin{document}

\begin{titlepage}
\rightline{LAPTH-697/98}
\rightline{UQAM-PHE-98/04}
\rightline{hep-ph/9809207}
\rightline{August 1998}

\vspace*{\fill}

\begin{center}

{\large{\bf Minimal Ten-parameter Hermitian Texture Zeroes Mass Matrices and the CKM
Matrix\footnote{Talk given by M.B. at the MRST 98, ``Towards the Theory of Everything", Montr\'eal,
13-15 May 1998.}}}

\vspace*{\fill}

{\bf M.~Baillargeon$^a$, F.~Boudjema$^b$, C.~Hamzaoui$^a$ and J.~Lindig$^c$ }

$^a$ D\'epartement de physique, Universit\'e du Qu\'ebec \`a Montr\'eal \\ C.P. 8888,
Succ. Centre Ville, Montr\'eal, Qu\'ebec, Canada, H3C 3P8 \\ $^b$ Laboratoire de Physique
Th\'eorique LAPTH \\  Chemin de Bellevue, B.P. 110, F-74941 Annecy-le-Vieux, Cedex,
France. \\ $^c$ Institute for Theoretical Physics, Leipzig University\\
 Augustusplatz 10, 04109 Leipzig, Germany

\vspace*{\fill}

\centerline{ {\bf Abstract} }
\vspace{12pt}
\baselineskip=18pt
\parbox{15cm}{
\noindent Hermitian mass matrices for the up and down quarks with texture zeroes but with
the minimum number of parameters, ten, are investigated. We show how these {\em minimum
parameter} forms can be obtained from a general set of hermitian matrices through weak
basis transformations. For the most simple forms we show that one can derive exact and
compact parametrizations of the CKM mixing matrix in terms of the elements of these mass
matrices (and the quark masses).}

\end{center}

\vspace*{\fill}

\end{titlepage}
\baselineskip=18pt

\section{Introduction}

\indent Within the three-family standard model, the Yukawa interaction provides ten
physical measurable parameters, the six quark masses and the four parameters of the CKM
mixing matrix\cite{CKM}. Although ten is certainly a large number if the model is to be viewed as
a fundamental theory, this number of parameters in fact emerges from two $3\times 3$
Yukawa matrices which in total amount to as many as 36 parameters! Clearly there is a
large number of redundant parameters. In order to better corner the mechanism of symmetry
breaking it could be advantageous to work with mass matrices that exhibit only the least
number of parameters, {\it i.e.} ten. Having to deal with a minimal set not only eases
the computational task, like going from the mass matrices to the mixing matrix, but when
confronting the set with the data this may even help in better exhibiting patterns of
mass matrices that hint towards further relations between some of the elements of the
set. In this case one can entertain the existence of symmetries and models beyond the
standard model that can explain the approximate relations. At the same time this general
minimum approach could also reveal whether some relations, like those that relate some
mixing angles to ratios of masses, are in fact not specific to a particular constrained
model but are rather generic in a much wider class of models.\\

 Most of the
{\em constrained} matrices that aim at relating the mixing angles to the quark masses,
and hence reduce the number of parameters to much less than the needed 10, are based on
so called texture zeroes mass matrices \cite{RRR}. These are based either on some specific
"beyond the standard model" scenario or by postulating some ad-hoc ansatz. In ans\"atze
were these zeroes are not related to any symmetry only the non-zero elements are counted
as parameters, although it is clear that there are numerous ways of keeping to ten
independent parameters. For instance instead of zeroes one can take some elements to be
equal or have any other definite relations between them \cite{democratic,usy}.
Democratic \cite{democratic} mass matrices with all elements equal are a case in mind,
these are a one-parameter model which when written in an appropriate basis can be turned
into matrices with all elements but one being zero. In our approach we keep within the
popular textures zeroes paradigm and look for those bases were only the minimum number of
(non-zero) independent parameters appears explicitly.

Obviously, dealing with less than ten parameters invariably leads to relations between
masses and mixings. In many cases and for a certain range of masses the less-than-ten
parameter descriptions may turn out not to be supported by data if the textures are
over-predictive. On the other hand if one works with, at least, ten independent
parameters then one should always reproduce the data since there should be possible to
make a one-to-one mapping. A general classification of symmetric/hermitian textures
zeroes with a number of parameters less than ten has been given in \cite{RRR}, while
Branco Lavoura and Mota \cite{BrancoMota} (BLM) have been the first to point out that for
non hermitian matrices some textures zero \`a la Fritzsch \cite{Fritzsch1} were just a
rewriting of the mass matrices in a special basis and thus the zeroes of the much
celebrated Fritzsch ansatz were `` void" of any physical content. In fact the BLM
approach for non hermitian matrices still involves twelve parameters, the extra two being
related to the phase conventions taken for the CKM\footnote{One should be fair and say
that, sometimes, keeping one or two of the redundant parameters may prove useful. However
we will stick with the minimalist description.}.

Recently an approach based on BLM has been pursued by some
authors \cite{Falcone,Koide,Harayama}, taking a specific pattern of non hermitian matrices
and in some cases re-expressing one of the mass matrices with the help of the
phenomenological parametrization of the CKM matrix. In this talk we will concentrate on
hermitian texture zero matrices having zeroes in the non-diagonal entries. The case with
zeroes on the diagonal will be presented in a longer communication \cite{ourpapermass}. Note that we
differ from \cite{RRR} in that we still have ten parameters. It is known that within the
standard model one can always express the mass matrices in a basis were they are
hermitian \cite{Frampton,BrancoMota,Mahermitian}. Also, in the case of non hermitian matrices our
results should be understood as applying to the hermitian square matrices, $H=MM^\dagger$
. We supply a systematic list of all possible texture zeroes that contain the minimal set
of ten parameters and show how these textures can be reached from a general set of two
hermitian matrices for up and down quarks, through specific weak basis transformation
which we construct explicitly. Among all the patterns that we list, one shows a
particularly very simple and appealing structure which has a direct connection to the
Wolfenstein parametrization \cite{Wolfenstein}. For this we have been able to
analytically construct a compact exact formulae for the mixing matrix.

\section{Simple Texture Zeroes Quark Mass Matrices and the Choice of Basis}
The key observation as concerns the search of a suitable basis,
ideally one with the maximum number of zeroes, is that starting
from any set of matrices for the up and down quarks, the physics is
invariant if one performs a weak basis transformation on the
fields. In the case of the standard model, one can choose any
right-handed basis for both the up quark fields ($u_R$) and the
down quark fields ($d_R$), as well as any basis for the doublets
of left-handed fields ($Q_L$). All these bases are related to each
other through unitary transformations, $u_R \rightarrow V_u u_R \;
;\; d_R \rightarrow V_d d_R \;;\; Q_L \rightarrow U_L Q_L$.
Therefore all sets of mass matrices related to each other through

\begin{equation}
M^{\prime}_u=U^\dagger_L M_u V_u \;\;\;\;\;\; M^{\prime}_d=U^\dagger_L M_d V_d
\end{equation}

give rise to the same physics (same masses and mixing angles in the charged current).

For hermitian matrices this means that weak basis transformations involve only a single
unitary transformation, {\it i.e.}, $U_L=V_u=V_d=U$ and therefore one can use either the
set $M_u, M_d$ or the set $M'_u, M'_d$ with $M'_f=U^\dagger M_f U$. In the case of
hermitian matrices, one is starting with a set of 18 parameters and the task is to find a
unitary matrix $U$ which can absorb 8 redundant parameters. This should always be
possible since a $3 \times 3$ unitary matrix has nine real parameters, but since an
overall phase transformation $U=e^{i\phi} {\bf 1}$ does not affect weak bases
transformations, a unitary matrix provides the required number of variables to absorb the
redundant parameters.
\\

\subsection{Phase transformations}
 One special case of this type of unitary transformations
which always proves useful, even in the case of non-hermitian matrices, is the one
provided by unitary phase transformations $U_{ij}=e^{i\phi_i}\delta_{ij}$. Because a
global phase does not affect the transformation, we set $\phi_1=0$ without loss of
generality. This type of matrix is therefore a two-parameter matrix . Applying this type
of transformation on both $M_u, M_d$ one has the freedom to choose $\phi_{2,3}$ such that
two phases out of the six contained in the hermitian $M_u, M_d$ can be set to zero. The
only restriction is that one can not, in general, simultaneously remove the phases of
both $M_u(ij)$ and $M_d(ij)$ ({\it i.e.} for the same $(ij)$). In any case, two
parameters, or rather phases, out of the 18 can always be removed this way.

\subsection{The simple case of a basis where one matrix is diagonal}
 It is always possible to take
$U=U_d^D$ ($U_u^D$), that is the unitary matrix that diagonalises
the down (up) matrix. In these specific bases where one matrix is
diagonal, the other, non-diagonal matrix, will then have no zero
in general but 9 real parameters (of which 3 can be taken as
phases in the non-diagonal entries). Applying an extra phase
transformation removes two phases and therefore one does indeed
end up with 3 parameters in $M_d$ (the masses) and 7 in $M_u$
making up a total of ten which is the minimal number.\\
It is worth mentioning that similar bases (where one of the matrices is
diagonal) have been studied in the literature but for the case of
non-hermitian matrices \cite{Falcone,Koide,Harayama}. It is easy to
see that one can easily recover these bases. Indeed, one can apply
on our hermitian matrices, the following transformations: assuming
one is starting with a diagonal $M_u$ take $U=V_u$ as a phase
transformation or simply just the unit matrix, then it is always
possible to choose $V_d$ such that $M_d$ turns into a
non-hermitian matrix but with extra zeroes. We leave the proof and
a discussion of these kind of (diagonal) bases to our longer
communication \cite{ourpapermass}.

\subsection{Non trivial cases: Non diagonal matrices with no diagonal zero}
 In the
above simple case one had three zeroes\footnote{Since one is dealing with hermitian
matrices, the number of zeroes is that contained on one side of the diagonal.}. In fact
requiring that one maintains 10 parameters, and in the case of hermitian matrices where
the zeroes are set on the off-diagonal elements, three is the maximum number of zeroes.
The non trivial cases are when these three zeroes are shared between the up-quark and
down-quark matrices, that is one off-diagonal zero in one matrix and two off-diagonal
zeroes in the other. Indeed, having more than three off-diagonal zeroes, four say, one is
left with the six real parameters on the diagonals plus two complex numbers which reduce
to two real numbers after a phase transformation has been applied and thus leading to
only 8 real parameters. Therefore, by requiring off-diagonal zeroes the problem is rather
simple: one only has to combine a matrix with one off-diagonal zero with a matrix with
two off-diagonal zeroes. For each of these matrices there are three possibilities of
where to put the zero. All in all, one counts 18 such possibilities or patterns. These
are displayed in Table 1.

\begin{table}
\begin{center}
\begin{tabular}{|c|c|c||c|c|c|}
\hline & $M_u$ & $M_d$ & & $M_u$ & $M_d$ \\ \hline
 1& (1,2) & (1,3) and (2,3)&10 & (1,3) and (2,3) & (1,2)\\
2 & (1,2) & (1,2) and(2,3)&11&(1,2) and (2,3) & (1,2)\\ 3 & (1,2) & (1,2)
and(1,3)&12&(1,2) and (1,3) & (1,2)\\ 4 & (1,3) & (1,3) and (2,3)&13 & (1,3) and (2,3) &
(1,3) \\ 5 & (1,3) & (1,2) and (2,3)& 14 & (1,2) and (2,3) & (1,3) \\ 6 & (1,3) & (1,2)
and (1,3)&15 & (1,2) and (1,3) & (1,3) \\ 7 & (2,3) & (1,3) and (2,3)&16 & (1,3) and
(2,3) & (2,3)
\\ 8 & (2,3) & (1,2) and (2,3)&17   &  (1,2) and (2,3) &       (2,3)    \\
9 & (2,3) & (1,2) and (1,3)& 18 & (1,2) and (1,3) & (2,3) \\ \hline
\end{tabular}
\end{center}
\caption{\label{allforms}Location of the zeroes for the 18
different forms.}
\end{table}

All of these combinations can in fact be classified in only two distinct cases which can
not be obtained from each other by a simple relabeling of the axes. Denoting the two
arbitrary hermitian mass matrices in those bases by $M_u=A^{\prime}$ and
$M_d=B^{\prime}$, these cases are explicitly:
\begin{equation}
\label{case1} A^{\prime}=\left(\begin{array}{ccc} A^{\prime}_{11} & 0 & 0 \\ 0 &
A^{\prime}_{22} & A^{\prime}_{23} \\ 0 & A^{\prime*}_{23} & A^{\prime}_{33} \\
\end{array}\right),\; B^{\prime}=\left(\begin{array}{ccc} B^{\prime}_{11} & 0 &
B^{\prime}_{13} \\ 0 & B^{\prime}_{22} & B^{\prime}_{23} \\ B^{\prime*}_{13} &
B^{\prime*}_{23} & B^{\prime}_{33} \end{array}\right)\;\mbox{\rm \underline{case I}}
\end{equation}
and
\begin{equation}
\label{case2} A^{\prime}=\left(\begin{array}{ccc} A^{\prime}_{11} & 0 & 0 \\ 0 &
A^{\prime}_{22} & A^{\prime}_{23} \\ 0 & A^{\prime*}_{23} & A^{\prime}_{33}
\end{array}\right),\; B^{\prime}=\left(\begin{array}{ccc} B^{\prime}_{11} &
B^{\prime}_{12} & B^{\prime}_{13} \\ B^{\prime*}_{12} & B^{\prime}_{22} & 0 \\
B^{\prime*}_{13} & 0 & B^{\prime}_{33} \end{array}\right)\;\mbox{\rm \underline{case
II}}.
\end{equation}

where $A^{\prime}_{11}$ is an eigenvalue. Of course, one can exchange the role of
$A^{\prime}$ and $B^{\prime}$ so that $A^{\prime}=M_d$ and $B^{\prime}=M_u$.

To prove the existence of these bases and show how they are reached, it is easiest to
first move to the basis where $A$ is diagonal.

Denoting the eigenvalues of $A$ by $\lambda_i$, $(i=1,2,3)$ and
$A^{\prime}_{11}=\lambda_1$, we have in the eigenbasis of A
\begin{equation}
\label{diagonal}
A=\left(\begin{array}{ccc}
\lambda_1 & 0        & 0 \\
0      & \lambda_2   & 0 \\
0      & 0           & \lambda_3\end{array}\right),\;
B=\left(\begin{array}{ccc}
B_{11}    & B_{12}   & B_{13} \\
B^*_{12}  & B_{22}   & B_{23} \\
B^*_{13}  & B^*_{23} & B_{33} \end{array}\right).
\end{equation}

The unitary matrix which leads to the form for $A^{\prime}$ in
both Eq.~\ref{case1} (Case I) and Eq.~\ref{case2} (case II) is
simply
\begin{equation}
U = \left(\begin{array}{ccc}
1       &     0     &    0    \\
0       &     x_2   &    x_3  \\
0       &     y_2   &    y_3\end{array}\right)
\end{equation}
with the complex numbers $x_2,x_3,y_2,y_3$ subject to the orthonormality conditions. It
is then trivial to find the appropriate combinations of
$x_2,x_3,y_2,y_3$ that lead to either $B^{\prime}$ in the above two cases
\cite{ourpapermass}. For instance in the first case, requiring $B^{\prime}_{12}=x_2
B_{12} + x_3 B_{13}=0$ gives the appropriate $U$. All other cases with two off-diagonal
zeroes in one matrix and one in the other are treated in an analogous way. The proofs are
obtained from case I and case II just by relabeling the indices. Of course, the case
where the two quark mass matrices make up between them only two-zeroes, being much less
constrained, is always easier to construct.

\section{CKM matrices from off-diagonal texture zeroes hermitian matrices}

The advantage of texture zeroes matrices yet accommodating all the ten physical
parameters is that they allow to easily express the mixing matrix solely in term of the
elements describing the mass matrices. One could then work backward and use the hierarchy
observed in a particular parametrization of the CKM mixing matrix, together with the
hierarchy in the masses, to exhibit further correlations in the elements of the mass
matrices expressed in a simple basis that already exhibits zeroes. \\

Recently, Ra\v sin \cite{Rasin} has devised a procedure to express
the CKM matrix as a function of the mass matrices in the general
case where no zero element is found in neither $M_u$ nor $M_d$. He
uses a product of rotation matrices and phase matrices to
diagonalise a general $3\times 3$ matrix. However, even when we
require $M_u$ to be diagonal, which is a special case
of~\cite{Rasin}, we are still left with large formulae which
include sines and cosines of angles for which only the tangent is
explicitly known. These results do not give compact expressions
for the CKM matrix elements. Only when more zeroes are imposed do
the results simplify. Even with the simple textures that are
displayed in Table~\ref{allforms}, the recipe given in
\cite{Rasin} leads to tedious and complicated formulae
\cite{ourpapermass} which moreover come with an ambiguity in
determining the signs of the sines and cosines. We will show that,
with the textures that are displayed in Table~\ref{allforms},
there exists a more compact way of expressing the CKM that does
not make use of any sines or cosines but exhibits the masses and
the elements of the mass matrices explicitly.\\

Each combination in Table~\ref{allforms} will lead to a particular parametrization of
the mixing matrix. We concentrate on parametrization 14 not only to illustrate how the
diagonalisation of the matrices is carried out exactly, and hence how one expresses the
CKM, but also because it leads to a parametrization of the Kobayashi-Maskawa matrix
which is directly related to the Wolfenstein parametrization \cite{Wolfenstein}.

To achieve this, we first apply a weak basis phase transformation to the form 14, such
that the only remaining phase is located in the up quark matrix. Thus one is dealing with

\begin{equation}
\label{huhd14} M_u=\left(\begin{array}{ccc} u & 0 & y e^{i\phi} \\ 0 & \lambda_c & 0 \\ y
e^{-i\phi} & 0 & t
\end{array}\right),\;\;
M_d=\left(\begin{array}{ccc} d & x & 0 \\ x & s & z \\ 0 & z & b
\end{array}\right)
\end{equation}
 Note that this parametrization allows to have as input, at the level of the mass matrices,
the physical mass of the charm quark, $\lambda_c$. In what follows all physical masses
will be denoted by $\lambda_i$, the index $i$ being a flavour index.

These mass matrices are diagonalised through the following unitary matrices
\begin{equation}
U_u=\left(\begin{array}{ccc} \sqrt{\frac{u_t}{\lambda_{ut}}}& 0 &
\sqrt{\frac{u_u}{\lambda_{ut}}}e^{i\phi}\\ 0 & 1 & 0 \\
-\sqrt{\frac{u_u}{\lambda_{ut}}}e^{-i\phi}&0&\sqrt{\frac{u_t}{\lambda_{ut}}}
\end{array}\right),
U_d = \left(\begin{array}{ccc} \sqrt{\frac{b_d d_s d_b}{\Delta\lambda_{ds}\lambda_{db}}}
& \sqrt{\frac{d_d b_s d_b}{\Delta\lambda_{ds}\lambda_{sb}}} \sigma & \sqrt{\frac{d_d d_s
b_b}{\Delta\lambda_{db}\lambda_{sb}}}
\\
-\sqrt{\frac{d_d b_d}{\lambda_{ds}\lambda_{db}}} & \sqrt{\frac{d_s
b_s}{\lambda_{ds}\lambda_{sb}}} & \sqrt{\frac{d_b b_b}{\lambda_{db}\lambda_{sb}}}
\\
\sqrt{\frac{d_d b_s b_b}{\Delta\lambda_{ds}\lambda_{db}}} & -\sqrt{\frac{b_d d_s
b_b}{\Delta\lambda_{ds}\lambda_{sb}}} \sigma & \sqrt{\frac{b_d b_s
d_b}{\Delta\lambda_{db}\lambda_{sb}}}
\end{array}\right)
\end{equation}
Such that

\begin{equation}
U^\dagger_{u,d} \;M_{u,d}\; U_{u,d}=\left(\begin{array}{ccc} \lambda_{u,d}&0&0\\
0&\lambda_{c,s}&0\\ 0&0&\lambda_{t,b}
\end{array}\right)
\end{equation}

and
\begin{eqnarray}
x_i &=& |x - \lambda_i| \;\;\; ({\it e.g.} \;\;u_t=|u-\lambda_t|)\\ \lambda_{ij} &=&
|\lambda_i - \lambda_j| \\ \Delta &=& |b - d| \\ \sigma &=& \mbox{sign of}\; (b - d)
\end{eqnarray}

Expressing the diagonalising matrices, $U_u,U_d$, with the help of the physical masses
keeps the expressions of these matrices very compact.
As $V_{\rm CKM}= U_u^\dagger U_d$, we can now write the CKM matrix exactly:
\begin{eqnarray}
V_{us} & = & \frac{\sigma\left(\sqrt{u_t d_d b_s d_b} +
           \sqrt{u_u b_d d_s b_b} e^{i\phi}\right)}
           {\sqrt{\Delta \lambda_{ut} \lambda_{ds} \lambda_{sb}}} \\
V_{ub} & = & \frac{\sqrt{u_t d_d d_s b_b} -
           \sqrt{u_u b_d b_s d_b} e^{i\phi}}
           {\sqrt{\Delta \lambda_{ut} \lambda_{db} \lambda_{sb}}}
\label{vub} \\ V_{cd} & = & - \sqrt{\frac{d_d b_d}{\lambda_{ds}\lambda_{db}}} \label{vcd}
\\ V_{cb} & = & \sqrt{\frac{d_b b_b}{\lambda_{db}\lambda_{sb}}} \label{vcb} \\ V_{td} & =
& \frac{\left(\sqrt{u_u b_d d_s d_b} e^{-i\phi} +
           \sqrt{u_t d_d b_s b_b}\right)}
           {\sqrt{\Delta \lambda_{ut} \lambda_{ds} \lambda_{db}}} \\
V_{ts} & = & \frac{\sigma\left(\sqrt{u_u d_d b_s d_b} e^{-i\phi} -
           \sqrt{u_t b_d d_s b_b}\right)}
           {\sqrt{\Delta \lambda_{ut} \lambda_{ds} \lambda_{sb}}} \\
J & \equiv & \frac{\det\left[M_u,M_d\right]}{2i \lambda_{uc} \lambda_{ut} \lambda_{ct}
\lambda_{ds} \lambda_{db} \lambda_{sb}}
         = \frac{\sqrt{u_u u_t d_d d_s d_b b_d b_s b_b}}{\lambda_{ut} \lambda_{ds} \lambda_{db}
              \lambda_{sb}} \sin\phi
\end{eqnarray}

We see that these expressions are surprisingly simple given that they come from the mass
matrices. Moreover, contrary to some ans\"atze, this type of CKM matrix can always be
made to fit the data.

Nonetheless, we are now in a position to exploit the mass hierarchies. One can take $x_x$
as a small perturbation, which means that in fact $\lambda_f \simeq f$ where $f$ refers
to a diagonal element. In other words this assumption amounts to requiring that the
diagonal elements of the mass matrices deviate very little from their corresponding
eigenvalues. We then have from eq. \ref{vcd} and \ref{vcb},
\begin{eqnarray}
d_d & \simeq & |V_{cd}|^2 \lambda_s, \\ b_b & \simeq & |V_{cb}|^2 \lambda_b.
\end{eqnarray}

We also have from eq.\ref{vub}

\begin{eqnarray}
V_{ub} \simeq \frac{1}{\sqrt{\lambda_t}}\biggl( \sqrt{ \frac{d_d b_b}{\lambda_b} }-
\sqrt{u_u} \;e^{i\phi}\biggr) \simeq - \sqrt{ \frac{u_u}{\lambda_t} }\; e^{i\phi}
\end{eqnarray}

where we have made the additional assumption that the terms involving the down-quarks are
quadratic in the ``perturbation" $d_d \times b_b$ compared to the term originating from
the up quark matrix: $u_u$. This additional assumption is stronger than the previous ones
since it also compares the strengths of the off-diagonal elements of the up {\em and }
down quark matrices. In any case with these mild assumptions one can now trade $d_d,b_b,
u_u$, {\it i.e.} $d,b,u$ for the moduli of $V_{cd}, V_{cb}$ and $V_{ub}$ and physical
masses (up to some signs).

Taking into account the size of $d_d$, $b_b$ and $u_u$, we can now write
\begin{eqnarray}
\label{ourpara} V_{\rm CKM}& \simeq & \left(\begin{array}{ccc} V_{ud} &
\sqrt{\frac{d_d}{\lambda_s}} & - \sqrt{\frac{u_u}{\lambda_t}}e^{i\phi} \\
-\sqrt{\frac{d_d}{\lambda_s}} & V_{cs} & \sqrt{\frac{b_b}{\lambda_b}} \\ \sqrt{\frac{d_d
b_b}{\lambda_s \lambda_b}} + \sqrt{\frac{u_u}{\lambda_t}} e^{-i\phi} &
-\sqrt{\frac{b_b}{\lambda_b}} & V_{tb}
\end{array}\right), \\
J & = & \sqrt{\frac{u_u d_d b_b}{\lambda_s \lambda_b \lambda_t}} \sin\phi.
\end{eqnarray}

It is interesting to see that in this parametrization the
$V_{CKM}$ can be split into elements which originate either solely
from the down-quark sector\footnote{A similar observation has also
been made in \cite{HouWong}.} or the up-quark sector. To recover a
phenomenologically viable mixing matrix, one could thus
concentrate on each sector separately. Moreover this
parametrization is equivalent to the standard Wolfenstein
parametrization
\begin{equation}
V_W = \left(\begin{array}{ccc} 1 - \frac{1}{2}\lambda^2 & \lambda & \lambda^3 A (\rho - i
\eta) \\ -\lambda & 1 - \frac{1}{2}\lambda^2 & \lambda^2 A \\ \lambda^3 A (1 - \rho - i
\eta) & -\lambda^2 A & 1 - {\cal O}(\lambda^4)
\end{array}\right).
\end{equation}
with
\begin{eqnarray}
\lambda &=& \sqrt{\frac{d_d}{\lambda_s}} \nonumber\\ A &=& \frac{\lambda_s}{d_d}
\sqrt{\frac{b_b}{\lambda_b}} \nonumber\\ \rho &=& - \sqrt{\frac{u_u \lambda_s
\lambda_b}{\lambda_t d_d b_b}} \cos\phi \nonumber\\ \eta &=& \sqrt{\frac{u_u \lambda_s
\lambda_b}{\lambda_t d_d b_b}} \sin\phi
\end{eqnarray}

Asking for maximal CP violation \cite{maxcp} sets $\phi = \pi/2$
and leads to $\rho = 0$.

  From the form of the $V_{CKM}$ matrix it is now an easy matter to find phenomenologically
viable quark mass matrices. Most direct from our study is the {\em general} feature that
if $d=(M_d)_{11}=0$, then $d_d=\lambda_d=m_d$ and therefore one has the rather successful
prediction \cite{GattoSartori,Weinbergmass}: $V_{us}\simeq \lambda \simeq
\sqrt{\lambda_d/\lambda_s}=\sqrt{m_d/m_s}$. Moreover, introducing the perturbative
parameter $\epsilon \ll 1$ and with all other parameters of order 1, we may write the
hierarchical matrices:

\begin{equation}
\label{huhde} M_u=\lambda_t \left(\begin{array}{ccc} 0 & 0 & c \; \epsilon^3 \; e^{i\phi}
\\ 0 & \lambda_c/\lambda_t & 0 \\
 c \; \epsilon^3  \; e^{-i\phi} &     0      &    1
\end{array}\right),\;\;
M_d=\lambda_b \left(\begin{array}{ccc} 0 & a \; \epsilon^3 & 0 \\ a \; \epsilon^3 &
\epsilon^2 & b \; \epsilon^2 \\ 0 & b \; \epsilon^2 & 1
\end{array}\right)
\end{equation}

This leads to $V_{us}=\sqrt{m_d/m_s}=a \; \epsilon$ whereas
$|V_{ts}|=b\; \epsilon^2=(b/a^2)\; |V_{us}|^2$ with ($b,a \sim 1$).
Therefore, if one identifies $\lambda=a \epsilon$ then $A=b/a^2$.
We could ``adjust" $a,b$ (and $c$) to better fit the data. The
forms in Eq.~\ref{huhde} bear some resemblance to those presented
in \cite{HouWong}, but note that we arrive at these forms from a
rather  different approach.

Also in the down sector, the ansatz reproduces the correct ratio of masses. Note also
that with the ansatz for the up-quark, copied somehow on that of the down quark, we get:
$|V_{ub}|\simeq c \; \epsilon^3$.

\section{Conclusions}

We have shown that without any assumption on the mass matrices
apart from hermiticity, it is always possible to find a quark
basis such that 3 off-diagonal elements are vanishing, allowing to
diagonalise unambiguously the mass matrices and obtain the mixing
matrix. The case where either $M_u$ or $M_d$ is diagonal (and
therefore all the 6 vanishing elements are contained in one single
matrix) is of special interest but leads to lengthy formulae for
the CKM matrix entries. In all other cases, we arrived at compact
formulae for the mixing matrix. These compact formulae that
express without any approximation the $V_{CKM}$ matrix in terms of
the masses and other elements of the mass matrices can be compared
to popular parametrizations of the CKM matrix. The exact forms
that we find make it transparent which further assumptions one can
make ({\it i.e} more zeroes) to simplify the structure of the mass
matrices and yet be compatible with the data. We have given one
such example, and in passing we have shown how starting from the
general 10 parameter bases, the mere assumption of one extra zero
in $(M_d)_{11}$ gives the famous
relation\cite{GattoSartori,Weinbergmass} $V_{us}=\sqrt{m_d/m_s}$
which is seen then to be rather generic to a large class of models
and ans\"atze. From there one can add more constraints, for
example we have presented a new ansatz which can be made to fit
the data quite well.

\section{Acknowledgements}

The work of M.~B. is supported by La Fondation de l'Universit\'e du
Qu\'ebec \`a Montr\'eal and the work of C.~H. is supported in part by
N.S.E.R.C. of Canada.

\end{document}